\documentclass[%
 reprint,
 amsmath,amssymb,
 aps,
]{revtex4-2}
\usepackage{textcomp, gensymb}
\usepackage{babel}
\usepackage{graphicx}
\usepackage{dcolumn}
\usepackage{natbib}
\usepackage{bm}

\begin{document}

\preprint{APS/123-QED}

\title{Experimental demonstration of the combined arm- and cavity-locking system for LISA}

\author{Jobin Thomas Valliyakalayil}
\email{JobinThomasValliyakalayil@anu.edu.au}
\author{Andrew Wade} 
\author{David Rabeling}
\author{Jue Zhang}
\author{Daniel Shaddock}
\author{Kirk McKenzie}
\affiliation{
Centre for Gravitational Astrophysics, Australian National University
}%

\date{\today}

\begin{abstract}
Laser frequency noise suppression is a critical requirement for the Laser Interferometer Space Antenna (LISA) mission to detect gravitational waves. The baseline laser stabilization is achieved using cavity pre-stabilization and a post-processing technique called Time-Delay-Interferometry~(TDI). To enhance the margins for TDI, alternate laser locking schemes should be investigated. A novel stabilisation blending the excellent stability of the arm with the existing cavity reference has been shown theoretically to meet the first-generation TDI margins. This locking system was designed to be implemented as a firmware change and have minimal or no changes to the LISA hardware. This paper experimentally verifies the hybrid laser locking technique by utilizing two references - an optical cavity, and an interferometer with delay imparted using 10~km of optical fiber. The results indicate the viability of the combination of arm-cavity locking system for LISA. They show the key benefits envisioned by this technique; suppression of the cavity fluctuations by the arm sensor (by 21~dB in this demonstration) and reduction of Doppler pulling of the laser frequency, a key technical challenge for arm locking.
\end{abstract}

\maketitle


\section{\label{Introduction}Introduction}
The Laser Interferometer Space Antenna, or LISA, is a proposed Gravitational Wave (GW) detector comprising a constellation of three spacecraft that orbit the Sun in a triangular formation with 2.5 million kilometres of separation. The interferometer aims to measure displacement between spacecraft to detect GWs in the band of 0.1~mHz to 1~Hz with a sensitivity goal of less than 15~pm/$\sqrt{\textrm{Hz}}$ for each arm-link~\cite{LISAL32017, LISAPPA1998, Colpi2024}. From a free-running laser frequency noise of 30/f~kHz/$\sqrt{\textrm{Hz}}$, it requires 14 orders of laser frequency noise suppression to meet the LISA sensitivity. The laser stabilization approach is to use a combination of locking to a fixed length Ultra-Low Expansion~(ULE) glass cavity using Pound-Drever-Hall (PDH) technique~\cite{Drever1983} and Time-Delay-Interferometry~(TDI)~\cite{Tinto2003, Shaddock2004}; which synthesizes equal arm-length interferometer combinations. 

While both these techniques are robust, TDI remains an active area of research due to its complex noise couplings and challenging experimental validation~\cite{Bayle2019, Hartwig2021}. An alternate stabilization technique is arm locking, which has been studied extensively in various literature~\cite{Sheard2003, Sutton2008, McKenzie2009, Yu2014}, including experimental demonstrations~\cite{Sheard2005} to validate the locking scheme. However, these arm-locking schemes require additional changes to the LISA baseline to accommodate for Doppler pulling~\cite{Wand2009}. Doppler pulling is an unavoidable (but reducible) technical challenge where the arm sensor ramps the laser frequency whenever it senses a Doppler shift in the laser frequency~\cite{McKenzie2009, Yu2014}.

Recently, a novel stabilization scheme was proposed that locks to both the cavity and the interferometer arm simultaneously~\cite{Valliyakalayil2022}, and may only require a bitstream upload/software update. This scheme aims to provide enough margin for TDI to the point where the first-generation TDI would be sufficient for post-processing. The technique removes the need for any additional optical hardware for implementation and in lieu, require a scalable digital controller, and precise Doppler shift knowledge to ensure that the residual Doppler pulling does not pull the laser from the cavity resonance.

The purpose of this paper is to demonstrate the dual arm-cavity locking technique on a bench-top experiment. We set up the PDH sensor using an optical cavity, similar to the one proposed for LISA, with a linewidth of 184~kHz. To realise the arm interferometer, a Mach-Zehnder interferometer is constructed with 10~km optical fiber delay line, scaling the length down from 2.5 million km to the equivalent of 15~km free-space separation. This results in the null frequencies being shifted from multiples of 60~mHz to multiples of 20~kHz. 
We successfully locked the laser to both sensors using digital controllers at a high bandwidth of $\sim$150~kHz, which encompasses 7 multiples of the null frequency. The lasers stayed locked for over 15 hours and we verified the transfer function of the hybrid system with the analytical models. In the experiment, the arm sensor suppresses the cavity fluctuations up to 21~dB in the band between 40~Hz and 50~kHz while the cavity is dominant at higher frequencies, allowing for a stable control system that incorporates the null frequencies, and at lower frequencies, reducing the Doppler pulling effect. Simulated Doppler shifts were introduced into the arm sensor and are shown to be suppressed by the cavity sensor with the expected transient response and steady-state behaviour after 0.33~s. The results from this experiment mirror the ones in~\cite{Valliyakalayil2022} and is a proof-of-concept of the arm-cavity laser stabilization.
\begin{figure*}[ht!]
    \centering
    \includegraphics[height=11cm,width=18cm]{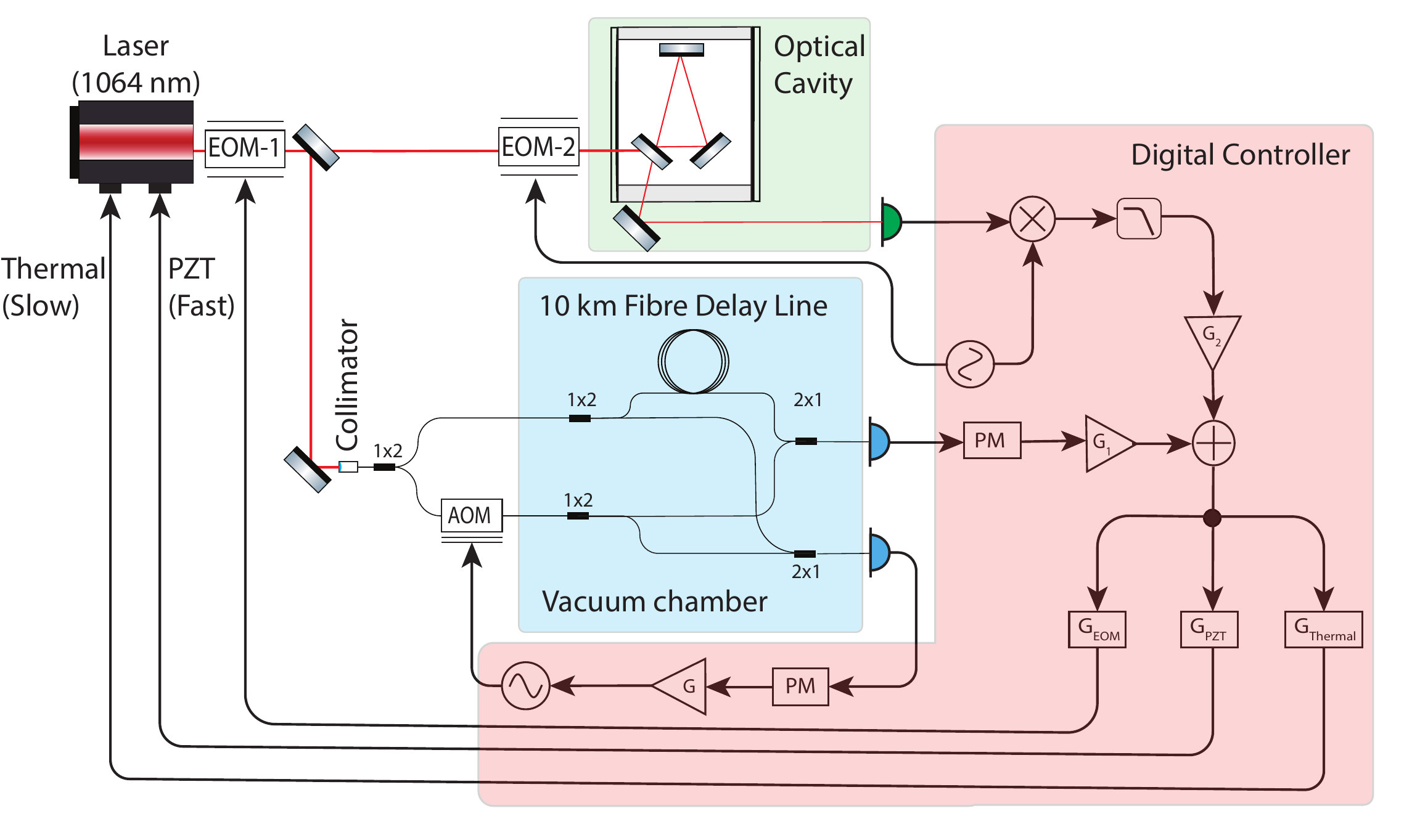}
    \caption{Layout of the arm-cavity locking experiment. The laser is passed through EOM-1, the actuation EOM, from where the closed loop performance is measured, and is split into both sensors. The PDH sensor is set up by passing the light to a modulation EOM, EOM-2, and then to the optical cavity, which provides the reflection on a photodetector. The arm sensor is constructed using a Mach-Zehnder interferometer with a 10~km delay line. An additional interferometer output is also taken to improve the noise floor of the arm reference by feeding back to the AOM. The photodetector output of each sensor is given to a digital controller, which blends both the sensor information and provides feedback to the three actuators using separate drivers. $G_1$ and $G_2$ represent the arm and the cavity controller, respectively, $G_{\textrm{EOM}}$, $G_{\textrm{PZT}}$ and $G_{\textrm{Thermal}}$ represents the different drivers for each actuator in the system, and PM refers to a digital phasemeter.}
    \label{Model1}
\end{figure*}

The paper is divided into six sections. Section~\ref{OpticalSetup} introduces the experimental layout used for the laser and the analytical models for each sensor. Section~\ref{DigitalSetup} describes the digital controller that is deployed for locking to each sensor, and the actuator drivers used for the feedback. Section~\ref{Doppler_pulling} explores the Doppler pulling effect in the experimental setup both at lock acquisition and steady-state behaviour. Section~\ref{Results} describes the key results from this experiment, while Section~\ref{Conclusions} provides the conclusions from the experiment and future implications for this stabilization technique.

\section{\label{OpticalSetup}Experimental Setup}
Figure~\ref{Model1} shows the experiment setup for locking the laser to both the delay line arm and the cavity references. The laser is a 200~mW Nd:YAG NPRO laser~(Lumentum Continous-Wave Single Frequency Infrared Series 126) operating at a wavelength of 1064~nm. We verified hthe free running laser frequency noise of this laser to be 3 kHz/$\sqrt{\textrm{Hz}}$ x (1/f) by comparing it to another cavity-stabilized laser known to have a noise of $<$ 1 Hz/$\sqrt{\textrm{Hz}}$ at 1 Hz. 

The laser has two actuators - 1. Fast actuation, using a Piezo-electric actuator made of lead-zirconate-titanate (PZT), and 2. Slow actuation, a thermal actuator that changes the temperature of the laser crystal. The fast actuator has a measured tuning range of 7.8~MHz/V with a bandwidth of up to 100~kHz, while the thermal actuator has a measured tuning of 14~GHz/V with a bandwidth of 0.1~Hz. In this experiment we scale the expected LISA arm lengths from 2.5 million km to a 15 km length achievable in an equivalent bench-top experiment; this necessitates scaling the bandwidth of the laser frequency actuation to encompass the higher null frequencies of the arm sensor. To achieve higher laser feedback control bandwidth in this experimnet, a fibre-coupled Electro-Optic-Modulator, or EOM~(iXblue), is used to extend the laser's frequency actuation bandwidth. A PZT is sufficient to reach the null frequencies for the longer arm lengths of the LISA mission itself. 

The light output from the actuation EOM~(EOM-1 in Figure~\ref{Model1}) is split into two, one for analyzing the closed loop performance, while the other path is split further for each sensor. The photodetector outputs from each sensor are digitized and provided to a commercial FPGA, a Moku:Pro, which combines the signals with frequency shaping and gives the appropriate output for each actuator. More details on the controllers are presented in Section~\ref{Controllers}

\subsection{\label{CavitySensor} Cavity sensor}
The cavity sensor uses a Ultra-Low Expansion glass (ULE) fixed resonator resonator based on a standard PDH locking setup~\cite{Drever1983}. A three-mirror travelling wave cavity is used with an effective round-trip length of 168.3~mm, yielding an FSR of 1.78~GHz. The linewidth is experimentally found to be 184~kHz, giving a finesse of the three mirrors to be around ~9680 similar to GRACE Follow-On~\cite{Folkner2010} and expected for LISA. 
The cavity assembly was mounted on an optical breadboard and is isolated by sorbathane dampeners (Thorlabs) and pneumatic vibration isolators (Newport) to provide isolation from vibrations.
\begin{figure}[h!]
    \includegraphics[width=8.5cm]{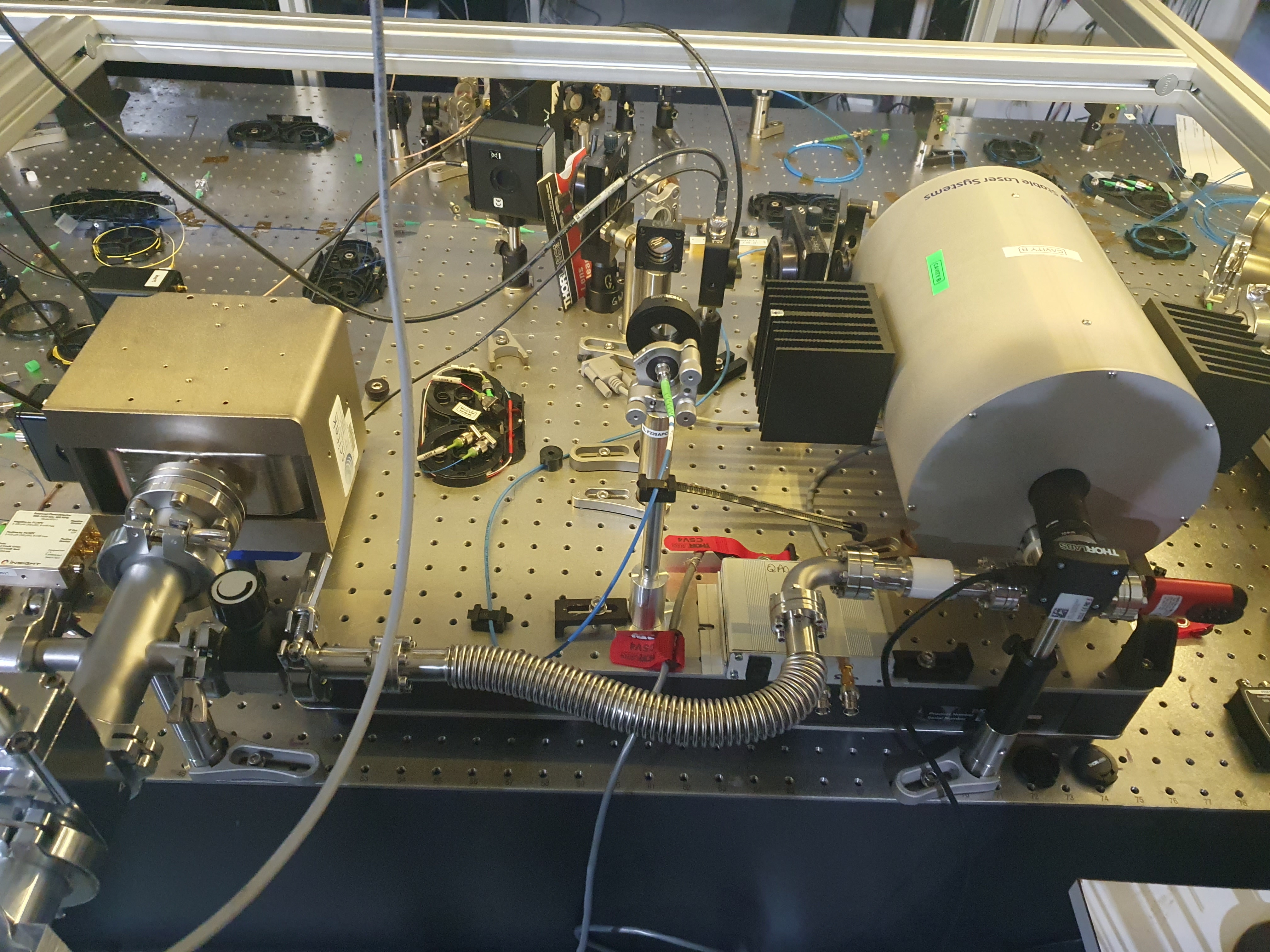}
    \caption{Experimental setup of the optical cavity. The triangular cavity is placed inside the vacuum chamber shown on the right side. The chamber is pumped through a roughing pump and is maintained at low pressure with an ion pump shown on the left-side.}
    \label{5CavitySystem}
\end{figure}

The laser is modulated at 12.259~MHz using a modulation EOM (identical but different from the actuator EOM). The reflected signal is detected on a 150~MHz bandwidth InGaAs Free-Space photodetector (Thorlabs) and after digital demodulation, the PDH error signal was extracted. For modelling the control systems, the sensor, $P_{\textrm{PDH}}(s)$, can be approximated as a low-pass filter~\cite{Hall2017}:
\begin{equation}
    P_{\textrm{PDH}}(s) = \frac{D_0}{1 + \frac{s}{2\pi~f_c}}
\end{equation}
Here `s' is the Laplace operator, and $D_0$ is the aggregate gain due to the cavity line-width, laser power, modulation depth and electronic gains. We estimate the value of $D_0$ to be 217 nV/Hz, while $f_c$ is the Half-Width-Half-Max~(HWHM) frequency of the optical resonator estimated to be 92~kHz.

\subsection{\label{ArmSensor} Arm sensor}
To mimic the arm interferometer for LISA, a Mach-Zehnder interferometer is set up using an optical fiber with a length of 10~km. A local oscillator~(LO) path is shifted by a beat note frequency ($\sim$62.5 MHz) using an Acoustic-Optic Modulator(AOM) (Free-space Isomet 1205C-843). The frequency-shifted light is passed to an isolation chamber through hermetically sealed fiber feed-throughs. The signal path is also sent through fiber feed-through to the isolation chamber where the large delay line is housed. This fiber spool and a paddle controller, for polarization control, are single-mode~(SM) fibers in the signal path, while the remaining fiber components are polarization maintaining~(PM) fiber. The signal and the LO signals are interfered to produce the beat note which is detected using a 400~MHz-bandwidth Balanced photodetector (Insight BPD-1) for intensity noise rejection. The beat note is demodulated using a digital phasemeter to get the phase and frequency information for feeding back to the laser.

To attenuate the coupling of acoustic and thermal noise, the delay line is placed in a vacuum chamber, pumped down to 20~mbar pressure. The tank is placed on sorbathane dampeners and on the same pneumatic isolated optical table as the cavity. The vibration noise coupling is further reduced by using a fiber spool encased in resin-like material. Inside the chamber, the spool is placed in a stainless steel thermal shield for further ambient temperature variation isolation. To achieve common mode rejection of the AOM noise and fibre drift outside the chamber, we utilise another interferometer without any delay to track and null these noise sources. This common-path noise cancellation ( $<$ 10~Hz) is done by feeding back to the AOM using the non-delay-line beat note with 10~kHz bandwidth and is detailed in Appendix~\ref{AOMCancellation}.

A length of 10~km on optical fibre translates to roughly 15~km of free-space delay accounting for the refractive index of glass. Thus, the arm sensor can be written as:
\begin{equation}
    \begin{split}
        P_{\textrm{arm}}(s) & = 1 - e^{-s\tau}= 1 - e^{-sL_{\textrm{arm}}n/c} \\
                & = 1 - e^{-s ~50\mu s}
    \end{split}
\end{equation}
where $L_{\textrm{arm}}$ is the length of the spool, $n$ is the refractive index of glass estimated at 1.477, and $c$ is the speed of light in vacuum approximated at $3\times10^8$~m/s. The null frequencies are computed to be at multiples of 20~kHz (1/$\tau$), which necessitates the use of the EOM (a high-bandwidth actuator) to observe the nulls within the control bandwidth. 

\begin{figure}[h!]
    \includegraphics[width=8.5cm]{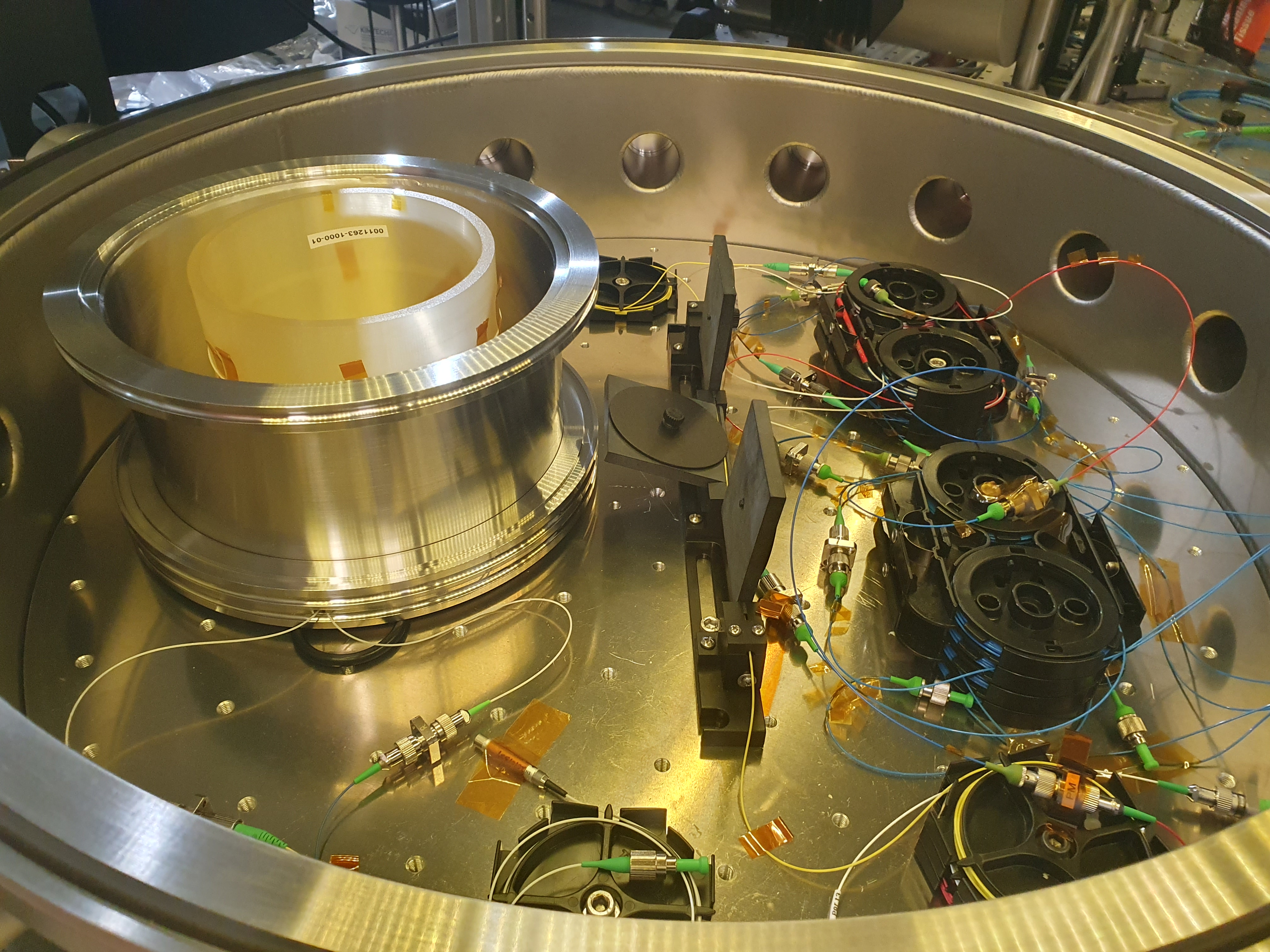}
    \caption{Experimental setup of the fiber Mach-Zehnder interferometer. The 10~km (50$\mu$s) arm delay line (left side) is placed in an isolated vacuum chamber pumped down to 2\% of atmosphere. Except for the spool and the paddle polarization controller, all the other components such as splitters/combiners and feedthroughs, wound in fiber trays (right side), are PM fibers.}
    \label{5ArmSystem}
\end{figure}

\begin{figure*}
    \centering
    \includegraphics[width=18cm]{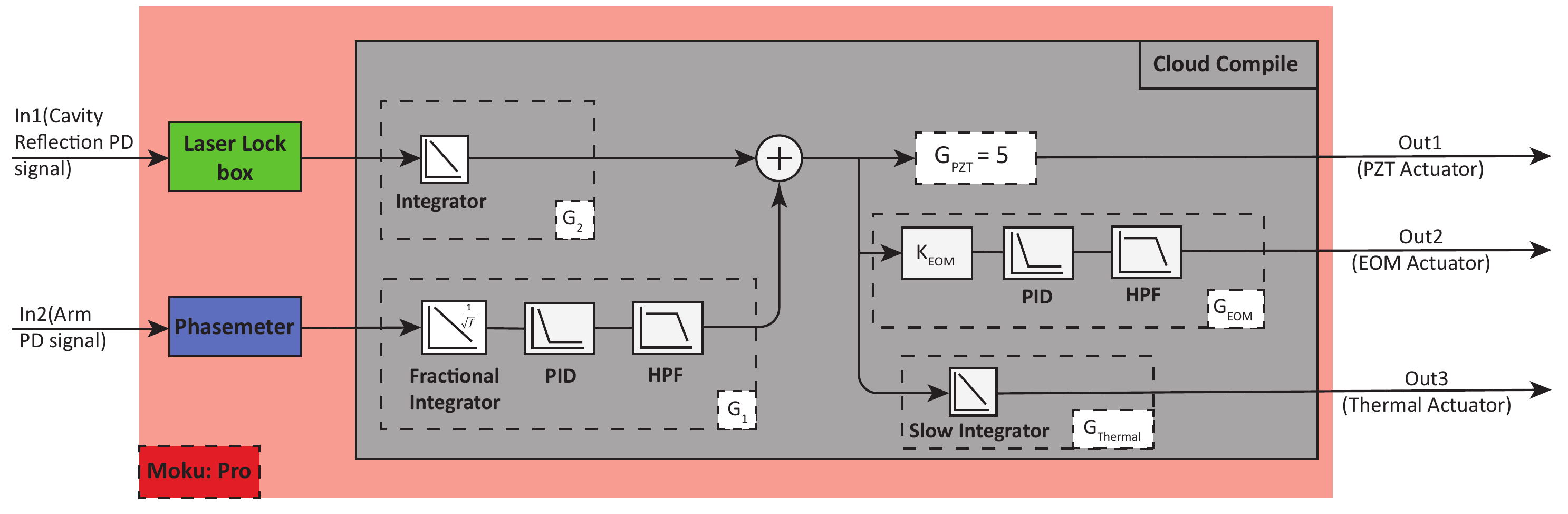}
    \caption{The Digital setup of the arm-cavity locking scheme on the Multi-Instrument Mode on Moku:Pro. The Laser Lock box and the phasemeter provide the cavity and arm sensor information, respectively, in the equivalent volts. The shaping controllers and actuator drivers are designed in Simulink and implemented using the Cloud compile service.}
    \label{MokuDigitalSetup}
\end{figure*}

\section{\label{DigitalSetup}Digital Controller Setup}
The two photodetector outputs from the two sensors, the PDH reflection output and the heterodyne interferometer beat note, are given to a commercial DSP platform on an FPGA~(Moku:Pro). The device is used in Multi-Instrument Mode to allow for parallel processing of the sensors. The PDH output is given to a Laser Lock Box instrument where the signal is demodulated at the PDH modulation frequency (12.259~MHz) and passed through a low-pass filter. The error signal is extracted from the filter output and scaled to equivalent volts for control. The error signal from the arm sensor is extracted using a phase-locked loop (PLL) phasemeter that measures the frequency deviations of the beat note frequency (62.5~MHz) and produces a voltage proportional to the frequency fluctuations. 

These two error outputs from each sensor are blended using a custom-coded digital controller for each sensor by shaping in frequency for stable loop operation. The frequency actuation signal is given to drivers (a combination of digital and analog electronics) that condition the signal to be used across each actuator for smooth cross-over. The digital controller and the actuator drivers are shown in Figure~\ref{MokuDigitalSetup}. The Cloud Compile labelled part in the Figure is designed in Simulink software and converted to a Hardware Descriptor Language~(HDL) using the in-built HDL coder by Mathworks~\cite{HDLCoder2023}. The bitstream is synthesized using the Cloud Compile service from Liquid Instruments and deployed on the Moku:Pro FPGA. 

\subsection{Controllers\label{Controllers}}

To ensure that the results can be extrapolated to those for the LISA mission~\cite{Valliyakalayil2022}, the controllers were re-designed to produce a similar band-sharing system. The shift of the null frequencies to high frequencies means the controller gains and slope were also changed. 
The arm controller, G$_1$, is split into three different parts, namely:

Stage I: Consists of an integrator whose slope is at -1.5. The steep slope of the integrator allows the cavity to dominate at high frequencies and provides more gain to the arm at low frequencies. The slope of the integrator is implemented as a sum cascade of low-pass filters and is detailed in Appendix~\ref{SlopeController}. \\
Stage II: Consists of a Proportional-Integrator (PI) controller to provide additional gain to the arm sensor.\\
Stage III: Consists of a 4th-order high-pass filter to ensure that the cavity is dominant again at lower frequencies ($<$ 30~Hz).

The combined arm controller is represented mathematically as:
\begin{equation}\label{G1}
    G_1(s) = \left({\frac{g_1}{s}}\right)^{1.5} \left({K_p + \frac{K_i}{s}}\right) \left({\frac{s}{s+p_{h_1}} \frac{s}{s+p_{h_2}}}\right)^2 
\end{equation}
The controller used for locking the cavity is an integrator whose unity gain frequency (UGF) is at 565.7~Hz.
\begin{equation}\label{G2}
    G_2(s) = \frac{g_2}{s}
\end{equation}

\begin{table}[h!]
\caption{Values for the gain, poles and zeros for the controllers as shown in Equations~\ref{G1} and~\ref{G2}}\label{ControllerValues}
\centering
\begin{tabular}{c c c}    
\hline\hline       
Parameter & Value & Description\\ 
\hline                    
   $g_1$ & 2$\pi$ $\times$ 0.17 & Gain of fractional integrator\\
   $g_2$ & 2$\pi$ $\times$ 565.69 & Gain of cavity integrator\\
   $p_{h_1}$ & 2$\pi$ $\times$ 300 rad/s & High-pass filter frequency-1 \\
   $p_{h_2}$ & 2$\pi$ $\times$ 3 rad/s & High-pass filter frequency-2 \\
   $K_p$ & 0.1965 & Proportional Gain in PI \\
   $K_i$ & 2$\pi \times$ 2.1713 $\times 10^3$ & Integrator Gain in PI\\
\hline                  
\end{tabular}
\end{table}

Any changes in the gain decide the band sharing between the two sensors, with these values being optimised to get the results in this paper. The optimization is found between the higher relative gain needed for the arm sensor to suppress laser frequency noise, traded off against the increased susceptibility to instability induced by the null frequencies of the delay line interferometer in the hybrid control. The controller outputs of each sensor are used for suppression function analysis in Section~\ref{Results}. The two controller outputs are summed together while maintaining data integrity/coherence in the two paths, and the blended signal with optimized gains and frequency crossover are applied to the laser light through several frequency actuators.

\subsection{\label{ActuatorDriver} Actuator Drivers}

To facilitate a stable feedback condition, a smooth cross-over between the actuators is requird with carefully chosen frequency cutoffs. In this experiment, the laser thermal actuation is dominant at low fourier frequencies, while the laser PZT is dominant from 0.1~Hz to 40~kHz, with the actuation EOM dominant from 40~kHz to high frequencies. These were accomplished through digital and analog electronic design. 

For additional gain/range to the PZT, the signal was boosted by a factor of five by a power amplifier.
\begin{equation}
    G_{\textrm{PZT;driver}}(s) = 5
\end{equation}
For the EOM, the driver consists of a gain scaling, two high-pass filters, and a PI controller. Mathematically, the driver's Laplace domain response is
\begin{equation}
    G_{\textrm{EOM;driver}}(s) = K_{\textrm{EOM}} \left({\frac{s}{s+p_{\textrm{EOM}}}}\right)^2 \left({1 + \frac{K_{\textrm{EOM};i}}{s}}\right)
\end{equation}
Here $K_{\textrm{EOM}}$ is a digital gain of 2500, $p_{\textrm{EOM}}$ is the high-pass filter frequency with a value of $2\pi\times1000$ rad/s, while $K_{EOM;i}$ has a value of $2\pi\times1.2\times10^4$. The high gain is required such that it prevails over the PZT at high frequencies, while the high-pass filters avoid saturation on the EOM output and thus provide a cross-over between the EOM and PZT outputs. Tuning these values distributes the actuator work between the PZT and the EOM; increasing the EOM gain makes the feedback system use the EOM at more frequencies than the PZT. Care must be taken to optimize between available EOM dynamic range and usable PZT bandwidth.

For the thermal actuator, the driver adds an extra integrator at low frequencies shown as:
\begin{equation}
    G_{\textrm{Thermal;driver}}(s) = \frac{2\pi\times8.834\times10^{-3}}{s}
\end{equation}
This output is attenuated by an analog 30~dB attenuator~(Mini-Circuits), allowing the thermal output to have comparable scaling to the PZT, and dominate at low frequencies with a single integrator at the cross-over.

\subsection{Total Controller response}
The complete open loop transfer function can be computed as:
\begin{equation}
\begin{split}
    G_{\textrm{open}} & =  \left({P_{\textrm{arm}}G_1 + P_{\textrm{PDH}}G_2}\right) \left({G_{\textrm{Thermal}} + G_{\textrm{PZT}} + G_{\textrm{EOM}}}\right)\\
    & = \left({P_{\textrm{arm}}G_1 + P_{\textrm{PDH}}G_2}\right) G_{\textrm{Actuator}}
    \end{split}
\end{equation}
where $G_{\textrm{Thermal}}, G_{\textrm{PZT}}$ and $G_{\textrm{EOM}}$ is inclusive of the model of the physical actuator and the drivers associated with them. The Laplace operator notation `(s)' is dropped for simplicity, for example, $G_{\textrm{open}}$ refers to $G_{\textrm{open}}(s)$.

\begin{figure*}[ht!]
    \centering
    \includegraphics[width=12cm, height=15.75   cm, angle = -90]{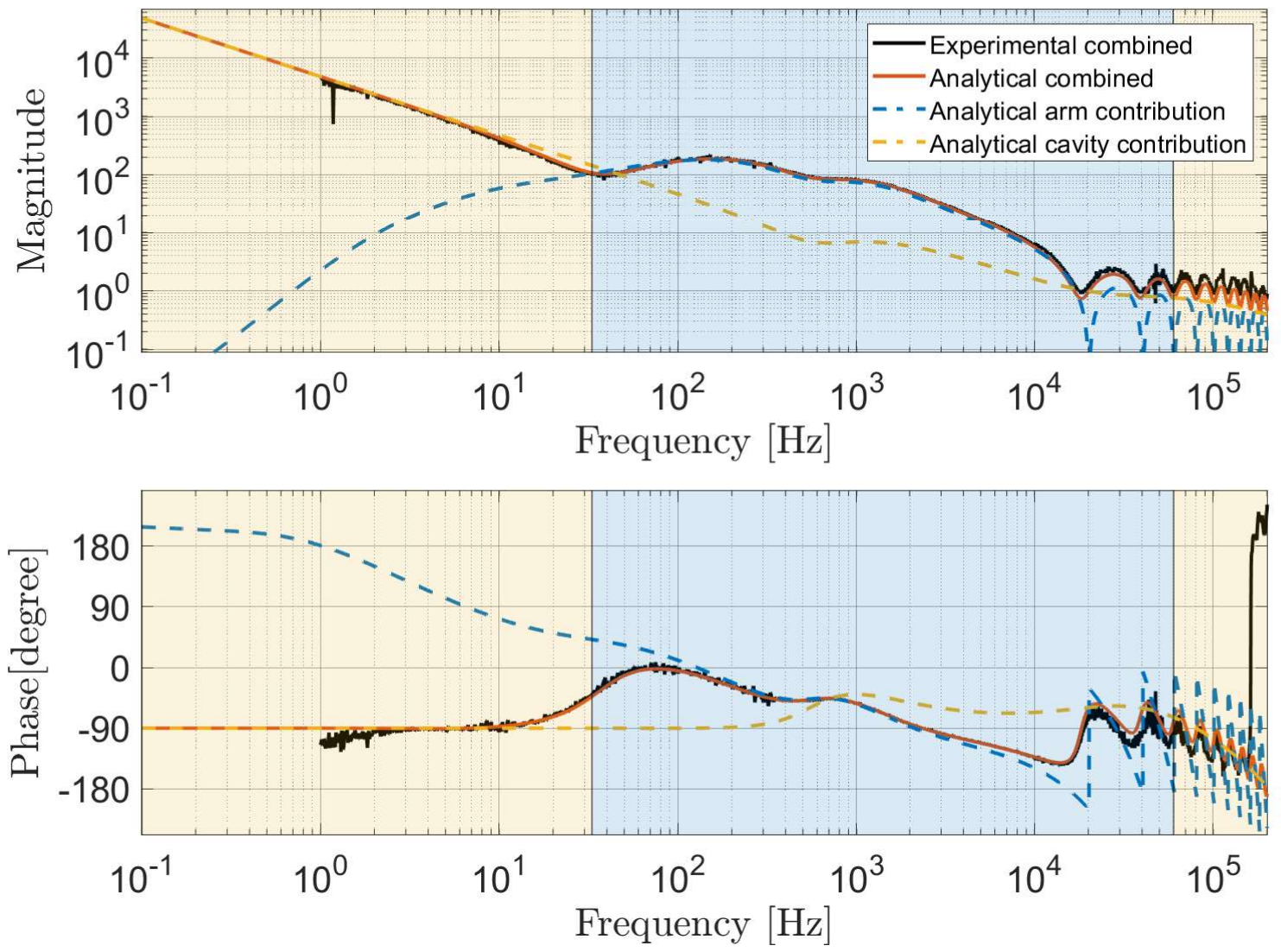}
    \caption{Experimental open loop bode plot of the hybrid control system. The black trace is the experimental open loop computed from the measured suppression of the laser frequency noise. The analytical traces shows a close match to the experimentally measured open loop transfer function, and using the same parameters, the individual contributions were computed analytically.}\label{EOpenLoop}
\end{figure*}

The total open loop transfer function of the experimental system is compared to an analytical model in Figure~\ref{EOpenLoop}. By measuring the total noise suppression of the control system, the total open loop transfer function is computed. From the data, the unity gain frequency of the combined controller is approximately 150~kHz, while encompassing 7 nulls within the bandwidth and having a maximum phase margin of greater than 20$\degree$. The arm is dominant from 40~Hz to 50~kHz, while the cavity is dominant in the remaining frequency bands. The combination of cavity with the arm sensor allows the control system to have a UGF at 150~kHz avoiding instabilities due to the phase variation of the nulls. The controllers designed here differ from those in~\cite{Valliyakalayil2022}, where the cavity controller is an integrator with a slope of 1.5, and the arm controller is an integrator with a slope of 2.3. These changes were made in the controllers to improve the visibility of the nulls at higher bandwidths without the phase delay causing instability.
Another difference is the high-pass filter order, which changes from a 7$^{\textrm{th}}$-order in~\cite{Valliyakalayil2022} to a 4$^{\textrm{th}}$-order in this experiment. The decreased filter order reduced the latency of the controller and increased the bandwidth of control.

A critical component when accommodating for the phase margin is the effect of physical delays introduced by the experimental setup that couple to the phase of the open loop as -2$\pi f_{\textrm{UGF}}\tau_{\textrm{delay}}$. The main contributions to the phase delay include the actuators, the controllers, and the ADC/DAC involved in providing inputs and outputs to the FPGA (Moku:Pro). The significant measured delays are shown in Table~\ref{DelayTable}, with other elements having negligible delays.
\begin{table}[h!]
    \centering
    \begin{tabular}{|c|c|}
        \hline
         Component& Delay ($\mu$s) \\
         \hline\hline
         PZT actuator & 3.7 \\
         EOM actuator & 0.6 \\
         Digital Controller & 0.57 \\
         ADC-DAC (combined) & 0.3 \\
         Bode Analyser (for suppression analysis) & 0.67 \\
         \hline
    \end{tabular}
    \caption{Physical delays of the key components in the feedback system.}\label{DelayTable}
\end{table}

The EOM and FPGA delays bound the maximum allowable bandwidth for the combined control system. A larger loop bandwidth is achievable with a different FPGA with lower delay; such a system would then be limited by the EOM. The phase margin of 30$\degree$ in~\cite{Valliyakalayil2022} considers the PZT to be the fast actuator but neglects the delay associated with the actuator and setup. A minor change to the cavity controller (dominant at the UGF), such as the addition of a lead compensator, will alleviate this problem.

Note that the implemented controllers and drivers are optimized to avoid internal quantization and saturation in the accumulators and filters. In the setup, the Stage I integrator of the arm controller was combined with one of the high-pass filters in Stage III, to achieve a single-order low-pass filter. Similarly, for the driver of the actuator EOM, the PI controller was combined with one of the high-pass filters to generate a proportional-low-pass filter controller. These changes avoided saturations in integrators if they were realized separately. The error signal from both the sensors has an in-built gain scaling function (in the Phasemeter and Laser Lock Box instrument) to convert to volts, and thus the overall gain of each controller can be distributed between the instruments and custom-designed controller.

\section{\label{Doppler_pulling}Doppler pulling}
\begin{figure}[h!]    
\centering
\includegraphics[width=7.5cm,height=10cm,angle=-90]{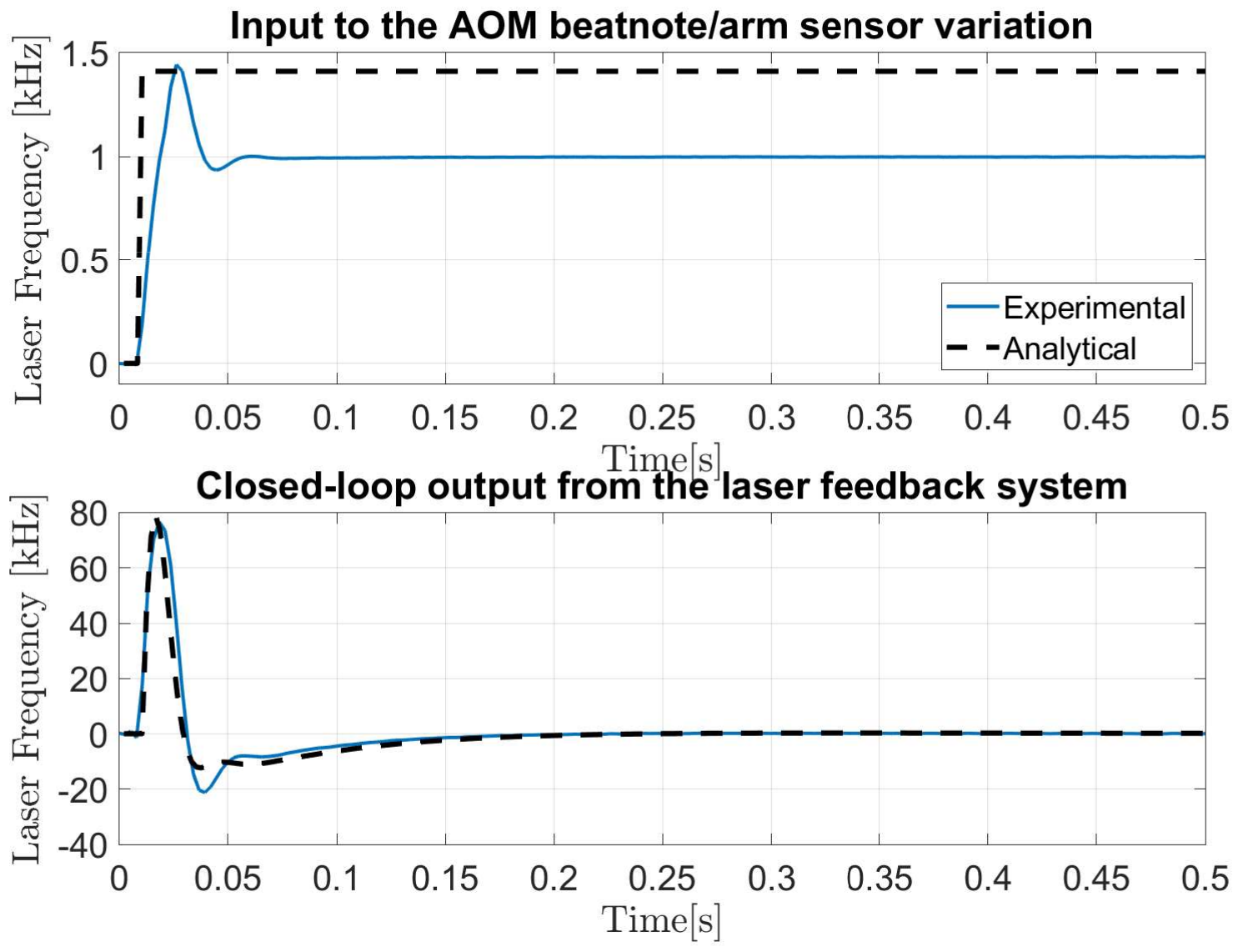}
    \caption{Doppler pulling for a step response to the AOM. The step given had an overshoot of 400~ Hz, which, when modelled into the system, shows a close match to the observed output. The resulting pulling is less than 80~kHz which was observed to be the maximum disturbance injected without breaking the cavity lock.}\label{EDopplerStep}
\end{figure}

To test the sensitivity of the experimental system to Doppler shifts, the RF tone to the AOM was modulated with step signals and sinusoidal tones to mimic Doppler shifts. The step response had an amplitude of 1~kHz, which had an overshoot of 400~Hz due to the low-pass response of the AOM and electronics giving an effective step size of 1.4~kHz. The experimental lock acquisition is shown in Figure~\ref{EDopplerStep}, with the step response recorded on the arm beat note. The analytical modelling assumes a perfect step function with an amplitude of 1.4~kHz. Both the experiment and the model showed an effective pulling up to 80~kHz before settling to a steady state after 0.33~s due to the high-pass filters in the arm controller. As the cavity HWHM frequency is 92~kHz, a higher frequency deviation of more than 1~kHz in the step will pull the laser out of the cavity resonance.

\begin{figure}[h!]
    \centering
    \includegraphics[width=7.5cm, height=10cm, angle = -90]{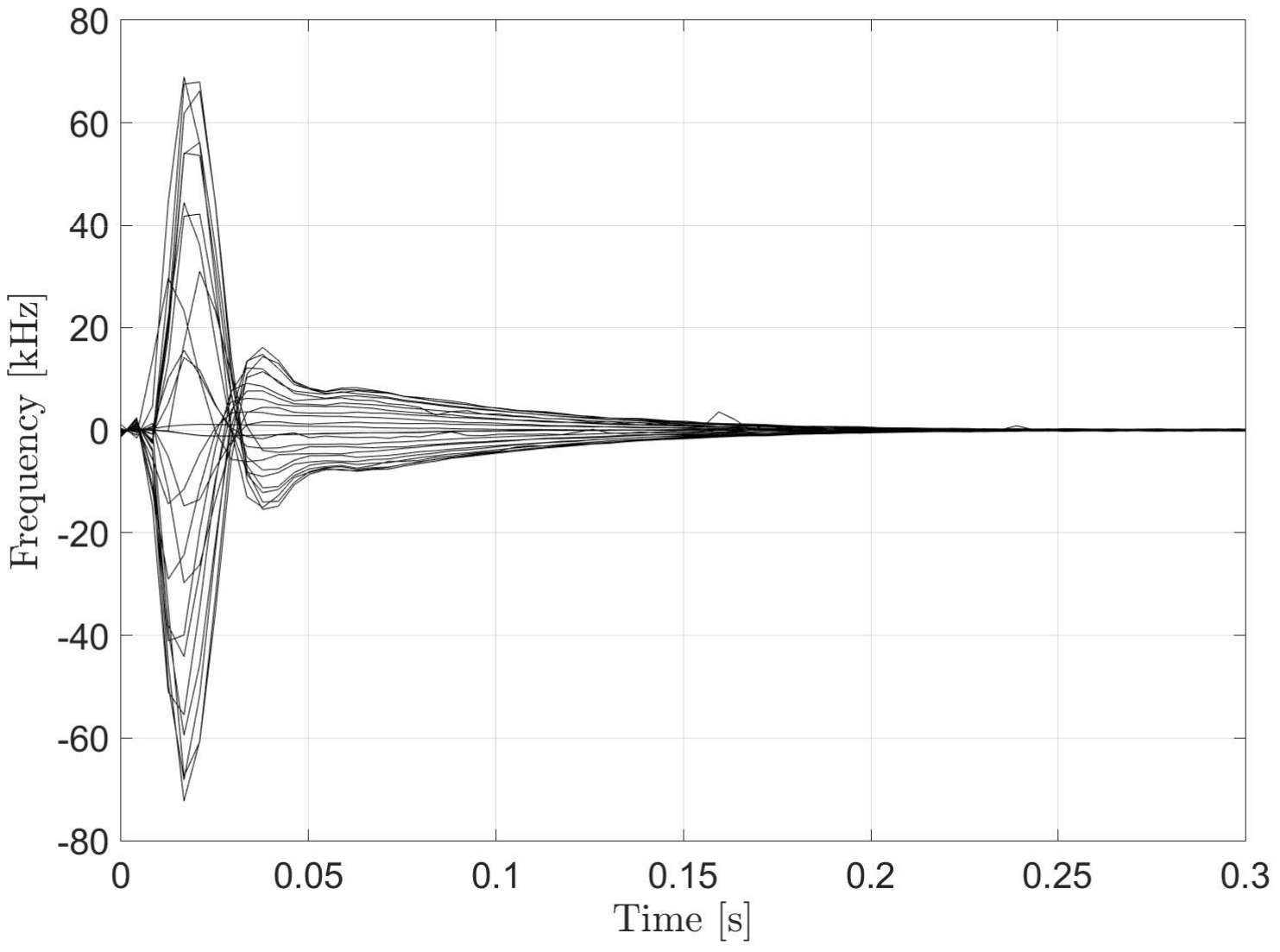}
    \caption{Doppler pulling at lock acquisition for sinusoidal tones at 0.3~Hz. The different traces correspond to the system's step response to the sinusoid's amplitude at different phase offsets. The resultant pulling is within the cavity resonance, maintaining simultaneous lock to the arm and cavity.}\label{EDopplerLockPhase}
\end{figure}

A similar test was also done using a sinusoidal tone at 0.3~Hz, with an amplitude of 1~kHz, while varying the initial phase of the sinusoid (from 0 to 2$\pi$) to illustrate locking at different set-points of Doppler bias. The different set-points initiate step responses of varying amplitudes based on the amplitude offset of the sinusoid at that set-point as shown in Figure~\ref{EDopplerLockPhase}. Like the step response, the lock acquisition of each trace settles after 0.33~s and reaches steady state behaviour. These results can be compared to Figure~5 of \cite{Valliyakalayil2022}, where the Doppler pulling is expected to go up to $\pm$~20~kHz which is less than the proposed cavity linewidth of 200~kHz and then settles after a period of 20 days.

\begin{figure}[h!]
    \centering
    \includegraphics[width=7.5cm, height=10cm, angle = -90]{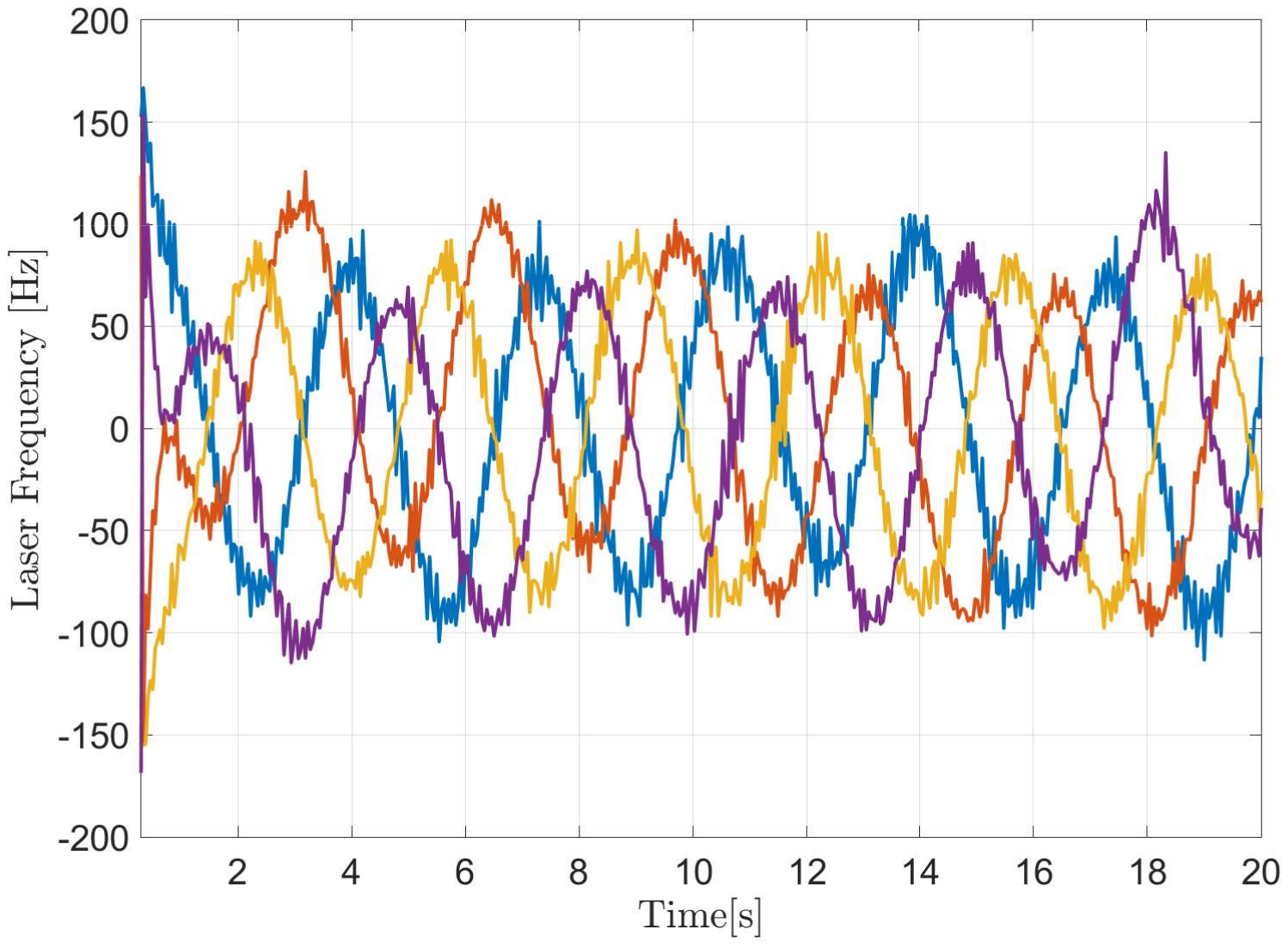}
    \caption{Doppler pulling in the steady state of the combined system. The injected disturbances are the same sinusoidal tone but with a 90$\degree$ difference between each consecutive one. Thus, each trace has a different lock acquisition behaviour, shown in Figure~\ref{EDopplerLockPhase}, but retains the same frequency with the amplitude suppressed by the cavity sensor. }\label{EDopplerSSPhase}
\end{figure}

The steady state behaviour of the Doppler pulling will depend on the suppression function of the arm noise by the cavity as shown in Figure~\ref{ENoiseSupp}. Figure~\ref{EDopplerSSPhase} shows the steady state response of key traces from Figure~\ref{EDopplerLockPhase} at 90$\degree$ phase offsets by observing beyond 0.33~s. The tone is observed at the same 
frequency of 0.3~Hz, with an amplitude roughly between 170-180~Hz-pp. For the tone at 0.3~Hz, the expected suppression is 21.1~dB, thereby reducing the amplitude of the sinusoid from 2~kHz-pp to 176.4~Hz-pp, matching with the experimental data. This is analogous to the steady state result of the work in~\cite{Valliyakalayil2022} that has a residual difference of $\pm$27~Hz. Due to the limiting noise floor of the experimental combined system, the minimum frequency tested was 0.3~Hz to observe the suppression of the tone at steady state. At lower frequencies, the suppression of the tone is less than the noise floor, and hence cannot be utilized to infer any residual Doppler pulling. Increasing the amplitude of the sinusoid can improve the visibility of the steady state but will also increase the chance of pulling the laser out of the cavity resonance at lock acquisition. For the LISA mission, the arm sensor is more stable and thus would have lower noise source than the Doppler shift residuals.

\section{\label{Results}Results}
\begin{figure}[h!]
    \centering
    \includegraphics[width=7.5cm,height=10cm,angle=-90]{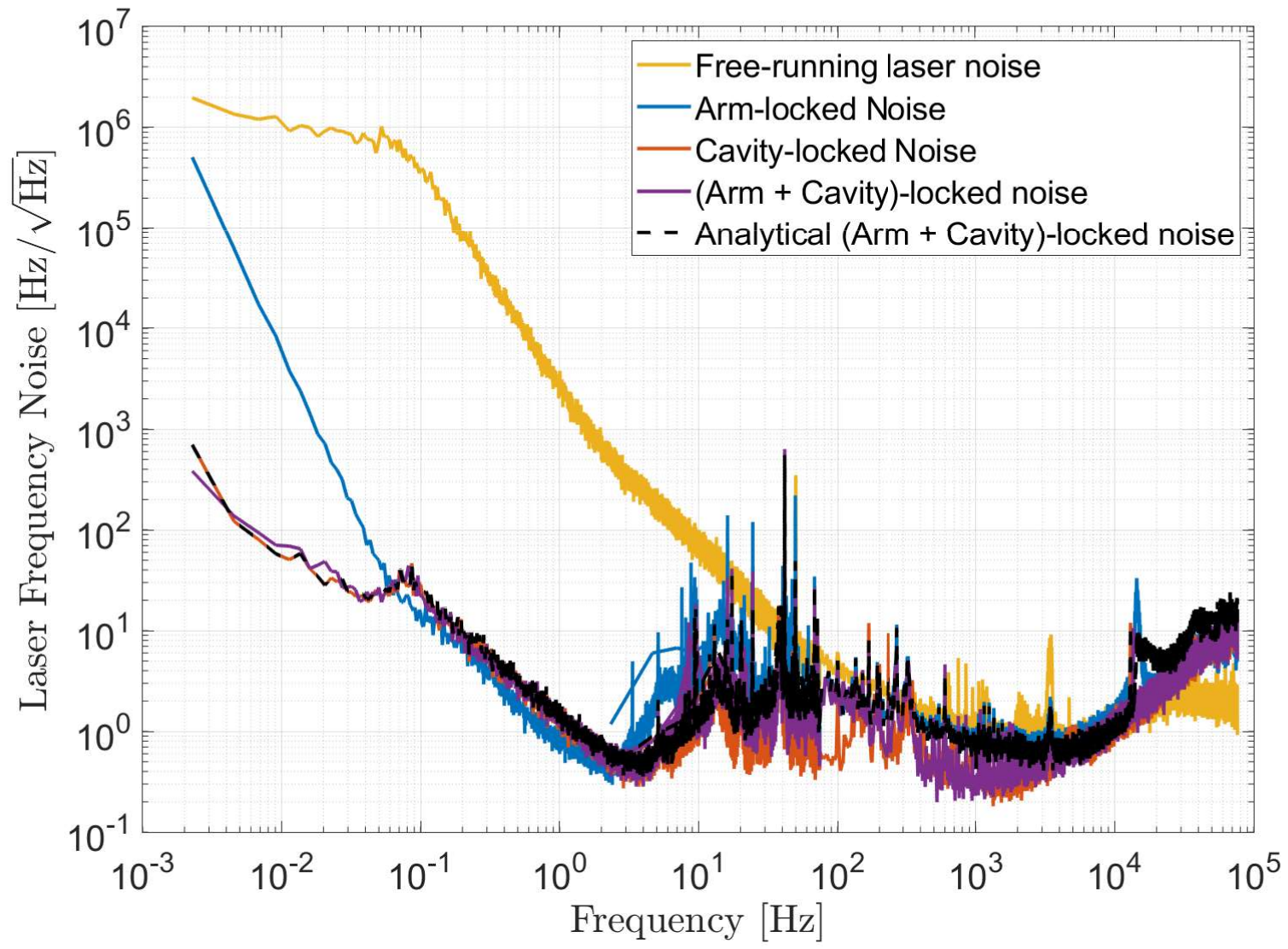}
    \caption{Noise spectrum of the free-running laser and laser that is stabilized by the arm interferometer, the optical cavity, and to both these references simultaneously. The noise floor of both references is similar at high frequencies ($>$ 0.1~Hz), while at low frequencies, the arm sensor has significantly higher thermally-induced frequency drift. }\label{ENoiseASD}
\end{figure}

Using the setup in~\ref{OpticalSetup}, and the controller in~\ref{DigitalSetup}, the noise spectrum is observed for three configurations -1. A standard PDH locked laser, 2. A Mach-Zehnder interferometer locked laser, and 3. A combined PDH cavity and Mach-Zehnder interferometer locked laser, and plotted in Figure~\ref{ENoiseASD}.
At low frequencies ($<$ 0.1 Hz), environmental thermal fluctuations dominate the noise in both sensors. The arm has more thermal drift with a coefficient of $\sim10^{-6}$/K, whereas the cavity is made of ULE glass with a temperature coefficient of $\sim10^{-8}$/K. At high frequencies ($>$ 5 kHz), the spectrum is limited by the control loop of the stabilized laser and that of the reference laser. In the intermediate frequency band, both the sensors have similar noise floors due to mechanical, vibrational, acoustic and air currents in the lab. These noise sources are common to both the sensors, as they are placed next to each other on the table, and thus, the arm or cavity dominant bands cannot be inferred directly from the noise spectra. The combined noise can be expressed using the individually measured noise spectra as:
\begin{equation}\label{5ExpArmCavityNSEqn}
\begin{split}
    \nu_C =& \frac{\nu_L}{1 + G_{\textrm{open}}} \\ & +\frac{\nu_{\textrm{Arm}}G_{1}G_{\textrm{actuator}}}{1 + G_{\textrm{open}}} + \frac{\nu_{\textrm{PDH}}G_{2}P_{\textrm{PDH}}G_{\textrm{actuator}}}{1 + G_{\textrm{open}}}\\
\end{split}
\end{equation}
where $\nu_C$ (purple trace in Figure~\ref{ENoiseASD}) is the closed loop laser frequency noise measured after the EOM-1 actuator, $\nu_L$ (yellow trace) is the free-running laser frequency noise,  $\nu_{\textrm{Arm}}$ (blue trace) is the noise coupling into the arm sensor, and $\nu_{\textrm{PDH}}$ (red trace) is the noise coupling into the cavity sensor.
\begin{figure}[h!]
    \includegraphics[width=7.5cm, height=10cm, angle = -90]{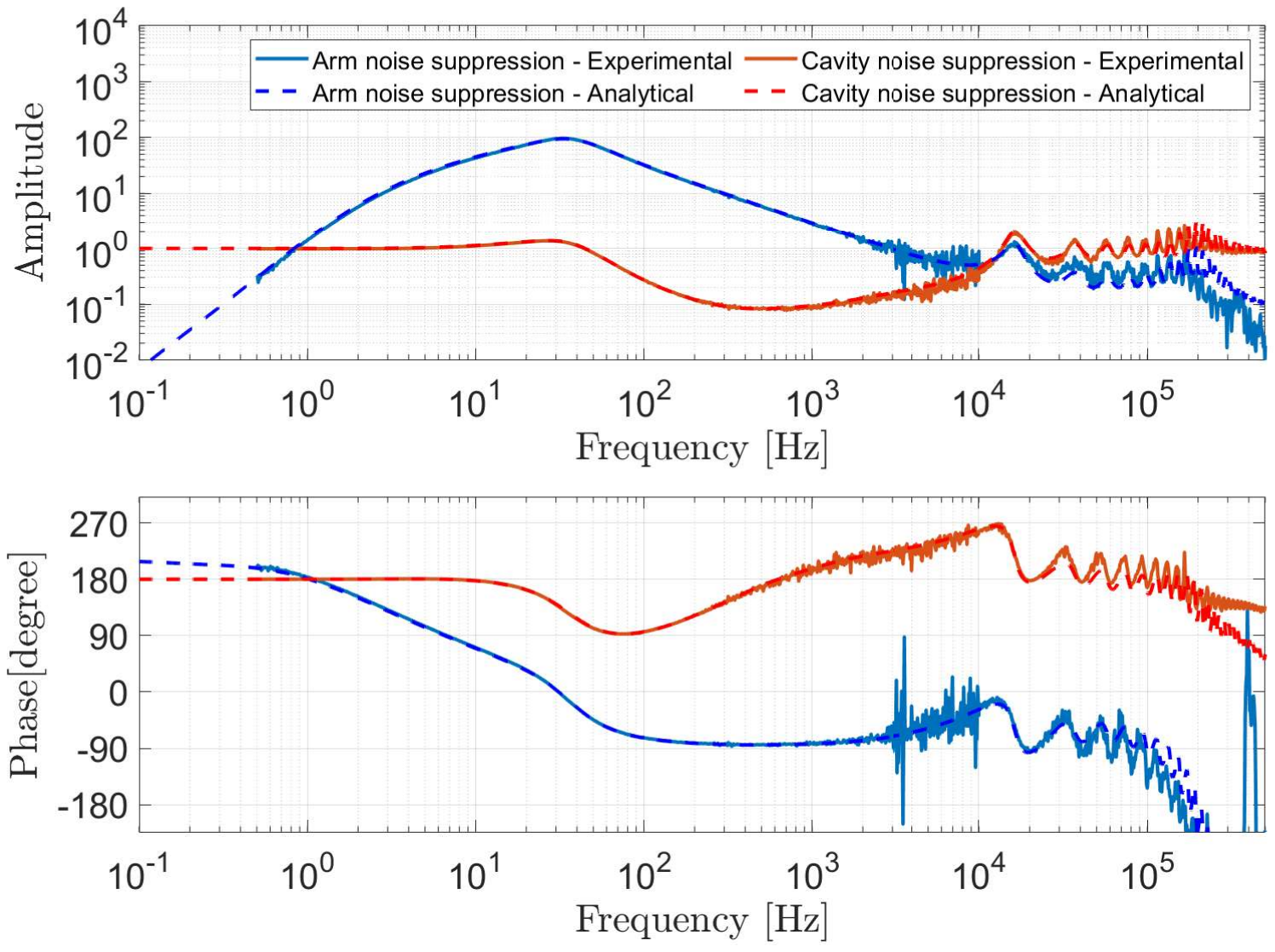}
    \caption{Suppression function of the individual sensor on the noise by the other sensor. The arm noise suppression by the cavity sensor allows sufficient reduction in Doppler pulling at low frequencies, while the cavity noise is suppressed by $>$~20~dB by the arm sensor. At 3-5 kHz, the transition between the actuators results in some low confidence in the transfer function.}\label{ENoiseSupp}
\end{figure}

To resolve the the frequency band where each sensor is dominant, analyses are done to obtain the suppression of the arm and cavity sensors individually. A sine signal was injected at the output of the arm sensor (phasemeter) as shown in Figure~\ref{BlockModel} by $\nu_{\textrm{inj1}}$, and the sum provided the cavity noise suppression by the arm, while the closed-loop laser frequency provided the arm noise suppression by the cavity (Refer to Appendix~\ref{BlockDiagrams}). Figure~\ref{ENoiseSupp} shows the suppression of arm noise at low ($<$ 1~Hz) and high frequencies (40~kHz), while the arm is able to suppress cavity fluctuations up to 21.5~dB. The experimental suppression function is compared with the analytical models based on Equation~\ref{5ExpArmCavityNSEqn}, and show a close match between the two. These models were also used to produce the analytical noise spectra in Figure~\ref{ENoiseASD} (black trace) using the other measured noise spectra, and the individual analytical traces in Figure~\ref{EOpenLoop}. The analytical and measured results match closely - validating the control system analysis for blending the two sensors and demonstrate the feasibility of implementing the same for the LISA mission itself.

\section{\label{Conclusions}Conclusion}
This paper presents the implementation and experimental results of an arm-cavity laser stabilization technique that was proposed for LISA in~\cite{Valliyakalayil2022}. By maintaining the same wavelength, 1064~nm, and a cavity with a small linewidth, 184~kHz, the experiment aims to closely match LISA parameters. The arm delay line to simulate the LISA arms was downscaled from 2.5 million kilometres to an effective free-space 15~kilometres using 10~km optical fiber. This is equivalent to shifting from 16.7~s to 50~$\mu$s or upscaling the null frequency to multiples of 60~mHz to 20~kHz. The arm sensor is based on single-arm locking rather than common arm locking in the original proposed scheme. However, this will have minimal effect on the analytical results. The control bandwidth was increased to 150~kHz, using an EOM as actuator, and allowed for the nulls of this down-scaled tabletop experiment to be visible within this bandwidth - a key requirement for validating the arm locking scheme for LISA.

The laser is individually stabilized to the optical cavity, and then to the interferometer arm using the hybrid controller to maintain locking to both references for more than 15 hours. From the controller design (using experimental and analytical approach), the overall UGF is at approximately 150~kHz, with the arm dominant from 40~Hz to 50~kHz, providing suppression of cavity fluctuations by up to 21.5~dB. The cavity is dominant in the remaining frequency band, and is important for reducing the Doppler pulling effect. The Doppler pulling is tested by injecting a step and sinusoidal tone  into the arm sensor. The amplitude of these disturbances are tested at the maximum before the cavity can lose lock, with the sinusoidal disturbances verifing the transient and steady state performance.
The noise suppression and Doppler pulling analyses verify that the arm-cavity locking in~\cite{Valliyakalayil2022} is a viable laser stabilization technique for LISA. The implementation of the scheme using a digital hybrid controller shows a flexible solution without any changes to the baseline optical setup proposed for LISA. 

\section*{Acknowledgment}
The authors acknowledge that this research was conducted with support from the Australian Research Council Centre of Excellence for Gravitational Wave Discovery
(OzGrav), through project number CE170100004.

\appendix
\section{Block diagrams and transfer functions}\label{BlockDiagrams}
The block diagram of the experimental arm-locking work is similar to the work in~\cite{Valliyakalayil2022}, with the main noise couplings shown as:
\begin{figure}[h!]
    \centering
    \includegraphics[width=9cm]{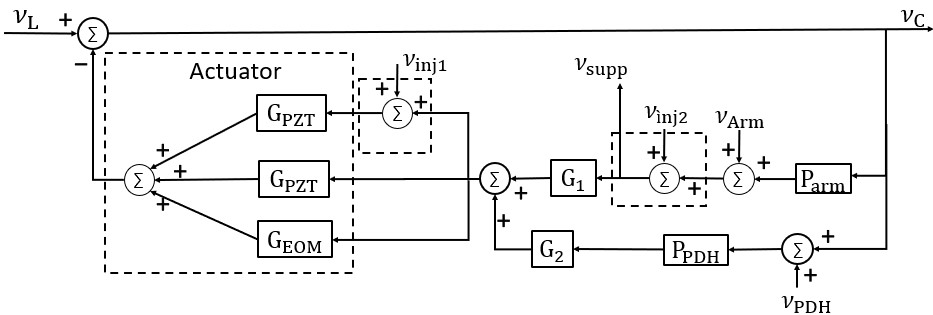}
    \caption{Block diagram of the hybrid arm-cavity locking system. }
    \label{BlockModel}
\end{figure}

Here, $\nu_L$ is the laser frequency noise, and $\nu_C$ is the closed loop performance measured after the EOM-1 actuator. The sensor noises, $\nu_{\textrm{Arm}}$ and $\nu_{\textrm{PDH}}$, represent the noise coupling in the arm interferomter, $P_{\textrm{Arm}}$ and the PDH readout for the optical cavity, $P_{\textrm{PDH}}$, respectively. 

The key transfer function analysis w.r.t Equation~\ref{5ExpArmCavityNSEqn} is obtained by injecting a tone at two spots - 1. After the sum of both sensor information and, 2. After the arm sensor information. Injecting a tone, $\nu_{\textrm{inj1}}$ allows us to see the total suppression function at $\nu_C$.
\begin{equation}
    \frac{\nu_{C}}{\nu_{\textrm{inj1}}} = \frac{G_{\textrm{PZT}}K}{1+G_{\textrm{open}}} \approx
    \frac{1}{1+G_{\textrm{open}}}
\end{equation}
Here K is an arbitary scaling factor and is chosen to be the inverse of the passband gain of the PZT transfer function, $\approx$ 130~nV/Hz. From this measurement, the open loop transfer function of the combination locking can be computed and is plotted in Figure~\ref{EOpenLoop}.

Injecting a tone, $\nu_{\textrm{inj2}}$ after the arm sensor allows us to see the noise suppression by each sensor on the other depending on the observed output. By looking at the summation right after injecting the tone, $\nu_{\textrm{supp}}$, shown in Equation~\ref{CavityNoiseSupp}, the suppression of the cavity noise by the arm sensor can be evaluated. Similarly, by observing the closed loop laser output, $\nu_{C}$, in Equation~\ref{ArmNoiseSupp}, the arm noise suppression by the cavity sensor can be computed. Both these suppression functions are plotted in Figure~\ref{ENoiseSupp}.
\begin{equation}\label{CavityNoiseSupp}
\begin{split}
    \frac{\nu_{\textrm{supp}}}{\nu_{\textrm{inj2}}} &  = \frac{1 + G_2 P_{\textrm{PDH}}}{1+G_{\textrm{open}}} = \frac{1}{1+G_{\textrm{open}}}  + \frac{G_2 P_{\textrm{PDH}}}{1+G_{\textrm{open}}} \\
    & \approx \frac{G_2P_{\textrm{PDH}}}{1+G_{\textrm{open}}} \hspace{1cm}\textrm{in arm dominant region}
    \end{split}
\end{equation}
\begin{equation}\label{ArmNoiseSupp}
    \frac{\nu_{\textrm{C}}}{\nu_{\textrm{inj2}}} = \frac{G_1}{1+G_{\textrm{open}}}
\end{equation}

 Equation~\ref{CavityNoiseSupp} is a variation of the actual cavity noise suppression shown in Figure 4 of~\cite{Valliyakalayil2022}. But as the total suppression function (first term) contributes less than the desired cavity noise (second term) suppression, the difference would not be significant in the arm-dominant region. For example, at 500~Hz where the maximum suppression is at -21.5 dB, the actual suppression would be -22.1dB computed from the model parameters. Due to limited input and output ports on the FPGA, the summation was done outside the digital platform using a custom-designed Opamps~(AD829AR) that added an extra delay of 0.67$\mu$s in the suppression function analysis.

\section{AOM beat note cancellation technique}\label{AOMCancellation}
In the experiment, the arm interferometer has a split before the isolation chamber, and thus a common-mode noise rejection scheme was sought to improve the noise performance of the fiber reference by recording two interferometer responses - 1. A prompt interferometer response which does not contain the delay line , 2. A delayed interferometer response which contains the optical fibre. If the time difference in the two interferometers are $\tau_{\textrm{Pr}}$ and $\tau_{\textrm{D}}$ respectively, the two arm responses can be shown as :
\begin{equation}
    P_{\textrm{Pr}} =  1 - e^{-s\tau_{\textrm{Pr}}} \hspace{1cm} P_{\textrm{D}} = 1 - e^{-s\tau_{\textrm{D}}}
\end{equation}
The subtraction between these two responses provides us a cancellation of the common-mode noise coupling into the interferometer outside the isolation chamber. Note that the laser noise would be dominant in the delayed interferometer arm response, compared to the prompt interferometer path, and thus arm locking will still be valid. Two methods were considered to implement this- 1. Directly subtracting the phase measurements of the two interferometers and use it for the laser feedback or 2. Using the prompt arm response to feed back to the AOM and use the delay interferometer for laser feedback.  For this work, the latter is used due to resource utilization of the Moku FPGA. The controller, G, is based on a simple integrator, and can be shown as :
\begin{equation}
    G = \frac{2\pi\times10000}{s}
\end{equation}
\begin{figure}
    \centering
    \includegraphics[width=9cm]{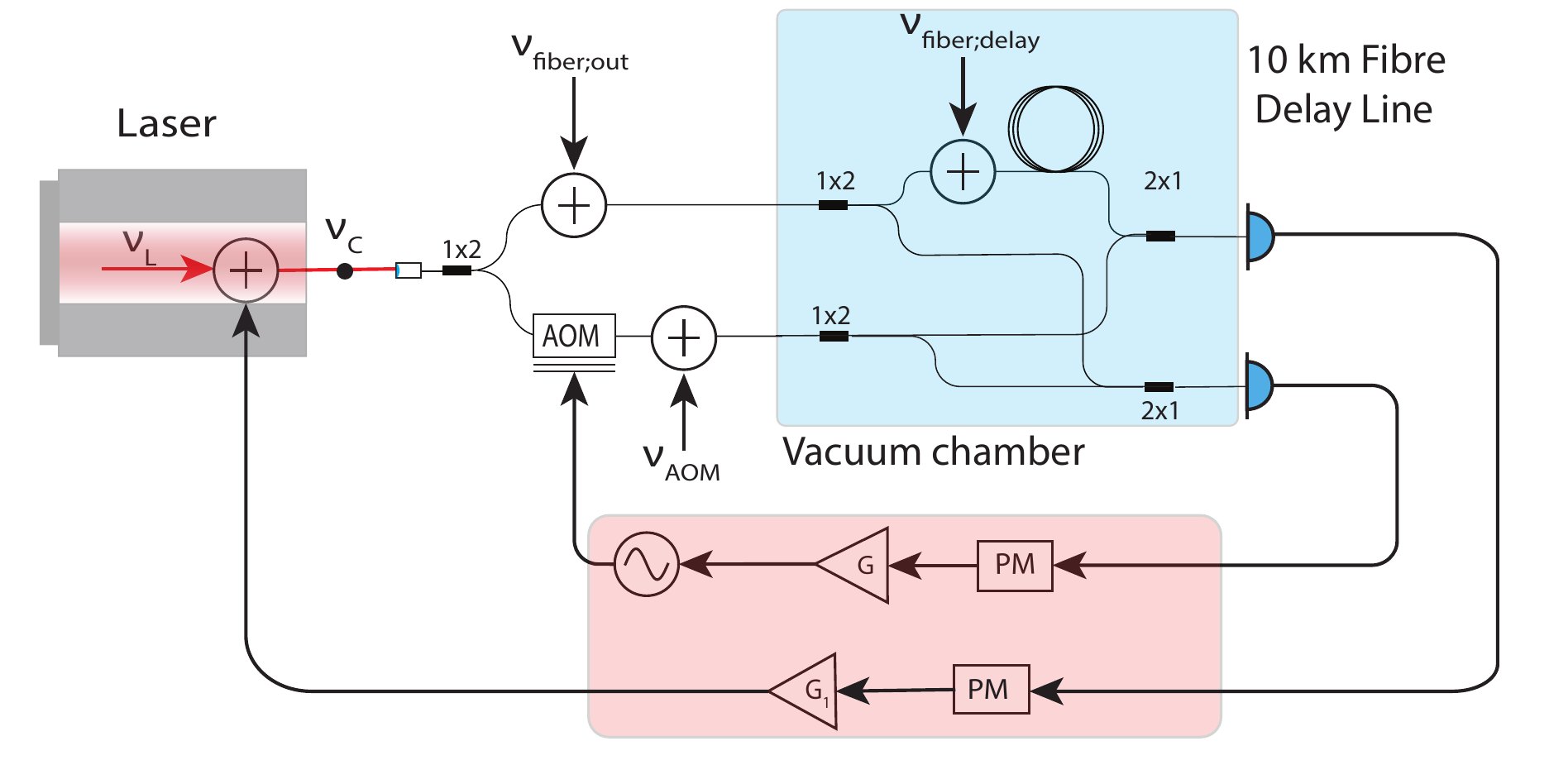}
    \caption{Model of the arm locking setup with the AOM cancellation technique.}
    \label{AOMBeatnoteModel}
\end{figure}
The noise sources considered here include, $\nu_L$, the laser frequency noise from the laser source, $\nu_{\textrm{fiber;delay}}$, is the fiber noise introduced by the delay line inside the isolation chamber, $\nu_{\textrm{AOM}}$ is the AOM-path related noise coupling, related to the LO path, while $\nu_{\textrm{fiber;out}}$ is the residual fiber noise outside the tank related to the Sig path. Here $G_1$ is the arm locking controller for locking the laser to the fiber reference. $\delta$ represents the cancellation efficiency of the phase subtraction used in the system, and is limited by the path-length difference in the LO-path for the two interferometers (the Sig part is not considered as the dominanting noise would be attributed to the optical fiber delay line in one path), and the phasemeter noise floor. Thus, the noise of the arm-locked laser can be simplified as:
\begin{equation}
     \begin{split}
        \nu_{C}  & = \frac{\nu_L}{1 + G_1(\frac{1}{1+G} + \frac{Ge^{-s\tau_{\textrm{Pr}}}}{1 + G}- e^{-s\tau_{\textrm{D}}})} \\ & + \frac{-\nu_{\textrm{AOM}}\frac{G_1\delta}{1+G} }{{1 + G_1 (\frac{\delta}{1 + G} + \frac{\delta~Ge^{-s\tau_{\textrm{Pr}}}}{1+G} - e^{-s\tau_{\textrm{D}}})}}  \\ & + \frac{\nu_{\textrm{fiber;delay}}G_1}{1+G_1\bigl( \frac{1}{1+G} + \frac{Ge^{-s\tau_{\textrm{Pr}}}}{1+G} - e^{-s\tau_{\textrm{D}}}\bigr)} \\ & + \frac{-\nu_{\textrm{fiber;out}}G_1(\frac{Ge^{-s\tau_{\textrm{Pr}}}}{1+G} - e^{-s\tau_{\textrm{D}}})}{1+G_1(\frac{1}{1+ G} + \frac{Ge^{-s\tau_{\textrm{Pr}}}}{1+G} - e^{-s\tau_{\textrm{D}}})}
     \end{split}
\end{equation}
Since $\tau_{\textrm{Pr}} << \tau_{\textrm{D}}$ by a factor of  $>$ 10000, $e^{-s\tau_{\textrm{Pr}}}$ can be approximated as 1 relative to $e^{-s\tau_{\textrm{D}}}$. Along with the high gain approximation of G, the output can be approximated as:
\begin{equation}
\begin{split}
    \nu_C & = \frac{\nu_L}{1+ G_1P_{\textrm{D}}} -\frac{\nu_{\textrm{AOM}}G_1\frac{\delta}{1+G}}{1+G_1(P_{\textrm{D}}+\frac{1}{1+G})} \\  & + \frac{\nu_{\textrm{fiber;delay}}G_1}{1+G_1P_{\textrm{D}}} - \frac{\nu_{\textrm{fiber;out}}G_1P_{\textrm{D}}}{1+G_1P_{\textrm{D}}}\\
    \end{split}
\end{equation}

\begin{figure}[h!]
    \hspace{-0.5cm}
    -\includegraphics[width=7.5cm,height=10cm,angle=-90]{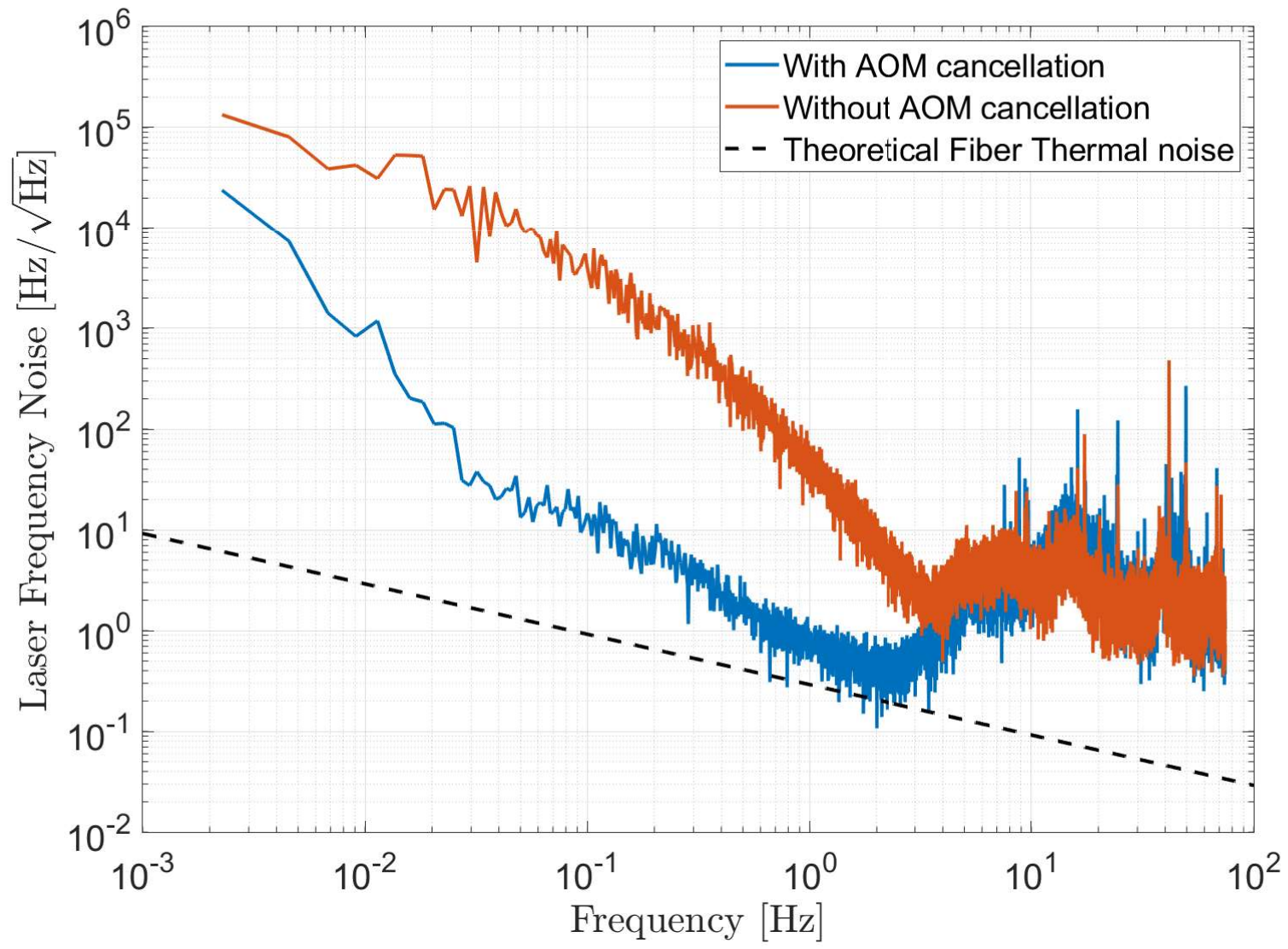}
    \caption{Noise spectra showcasing the effect of the AOM beat note cancellation on the stability of the arm reference. The additional beat note without the delay line helps in reducing the noise floor at low frequencies by a factor of 10 to 100. At 1-2 Hz, the noise spectra is close to the thermal limit of the optical fiber computed theoretically.}
    \label{5ArmNoisev1}
\end{figure}
The coupling of the laser phase noise, $\nu_L$, and the fiber noise in the delay line, $\nu_{\textrm{fiber;delay}}$, is expected and represent the stability of the arm reference with $P_\textrm{D}$ being the same as $P_\textrm{arm}$ from the analysis in this paper. The coupling of the AOM noise, $\nu_{\textrm{AOM}}$ can be shown to be reduced, and will be limited by the difference in the split paths of the LO path. The fiber noise emanating in the Sig path from outside the tank, $\nu_{\textrm{fiber;out}}$ is shown to be suppressed compared to the fiber noise that couples inside the tank, $\nu_{\textrm{fiber;delay}}$. This suppression arises from the arm response implicitly formed between the prompt and delay interferometer. For example, the transfer function scales the noise by $\frac{1}{2\pi\tau_{\textrm{D}}} \approx 3240$ without AOM feedback while it scales with unity with AOM feedback. The improvement of doing the AOM cancellation is shown in Figure~\ref{5ArmNoisev1} with a factor of 10 reduction at 1 Hz and factor of 100 at 0.1 Hz, with the resulting noise spectra almost reaching the theoretical fiber thermal limit at 1-2~Hz frequencies~\cite{Zhang2022}. 
\section{Controller for implementing 0.5 slope}\label{SlopeController}
The implementation of Stage I of the arm controller was achieved as a sum of multiple low-pass filters with appropriate gain as
\begin{equation}
    G_{1;I}(s)=\frac{g_0}{s}\sum_{i=1}^{10}\frac{g_i}{s+p_i}.
\end{equation}

\begin{table}[h]
\caption{Parameters of the gains and poles for the Low-pass filter cascade to implement $1/{s^{0.5}}$}\label{LP_table}      
\begin{tabular}{c c c}     
\hline\hline       
Index & Pole (p$_i$) & Gain(g$_i$) \\ 
number (i) & & \\
\hline           

  0 &   & (2$\pi$ x 0.162)$^{1.5}$\\
  1 & 2$\pi \times$ 0.1 & 1.878 \\
  2 & 2$\pi \times$ 1 &  4.591 \\
  3 & 2$\pi \times$ 10 & 14.375 \\
  4 & 2$\pi \times 10^{2}$ & 45.902 \\
  5 & 2$\pi \times 10^{3}$ & 1.4680 $\times 10^{2}$ \\
  6 & 2$\pi \times 10^{4}$ & 4.699 $\times 10^{2}$  \\
  7 & 2$\pi \times 10^{5}$ & 1.503 $\times 10^{3}$ \\
  8 & 2$\pi \times 10^{6}$ & 4.776 $\times 10^{3}$ \\
  9 & 2$\pi \times 10^{7}$ & 1.421 $\times 10^{4}$ \\
  10 &  0 & 1.571 \\

\hline                  
\end{tabular}
\end{table}

\end{document}